\documentclass[twocolumn,showpacs,preprintnumbers,amsmath,amssymb,floatfix]{revtex4}
\usepackage{epsfig}
\usepackage{dcolumn}
\usepackage{bm}

\begin{document}
\title{Renewal Aging as Emerging Property of Phase Synchronization}
\author{Simone Bianco$^1$ }
\author{Elvis Geneston$^{1}$}
\author{Paolo Grigolini$^{1,2,3}$}
\author{Massimiliano Ignaccolo$^{1}$}
\affiliation{$^1$Center for
Nonlinear Science, University of North Texas, P.O. Box 311427,
Denton, Texas 76203-1427, USA}
\affiliation{$^2$Istituto dei
Processi Chimico Fisici del CNR, Area della Ricerca di Pisa, Via
G. Moruzzi, 56124, Pisa, Italy}
\affiliation{$^3$Dipartimento di Fisica "E.Fermi" - Universit\'{a}
di Pisa, Largo
  Pontecorvo, 3 56127 PISA}
\date{\today}
\begin{abstract}
In this letter we examine a model recently proposed to produce phase synchronization [K. Wood \emph{et al}, Phys. Rev. Lett. {\bf 96}, 145701 (2006)] and we 
show that the onset to synchronization corresponds to the emergence of an intermittent process that is non-Poisson and renewal at the same time. We argue that 
this makes the model appropriate for the physics of blinking quantum dots, and the dynamics of human brain as well.
\end{abstract}


\maketitle

Phase synchronization of coupled clocks (oscillators) is a growing field of research, which is fast developing from the seminal work of Winfree
\cite{winfree} and Kuramoto \cite{kuramoto}. Using the model of coupled clocks of these authors it is possible to derive
\cite{biosa} the Turing structure \cite{turing}, which, in turn, triggered the research work in the field of diffusion
reaction \cite{prigo}. Physiologists are using a clock as a representation of a single neuron. In the work 
of \mbox{Ref.~\cite{stam}} a neuron is a clock whose behavior is described by a chaotic attractor: the R\"{o}ssler oscillator \mbox{\cite{rossler}}. 
The authors of \mbox{Ref.~\cite{geier}} and of \mbox{Ref.~\cite{katja}} have shown that coupled stochastic clocks 
can show cooperative (synchronized) behavior.
Brain functions, such as a cognitive act, rest on the cooperative behavior of a collection of many neurons \cite{varela}. 
The authors of \mbox{Ref.~\cite{eeg}} study the dynamic of the cooperation of a collection of many neurons by mapping the brain activity into a network. They 
find that the changes in the topology of the network describing the brain activity are driven by a non-Poisson renewal process operating in 
the non-ergodic regime \cite{nonergodic}. 
A collection of Blinking Quantum Dots (BQDs) \cite{bqd} has the same dynamical property as the brain activity.
The non-Poisson non-ergodic character of the distribution of the sojourn times of the BQDs in the ``light'' and in the ``dark'' state is
a well established property \cite{bqd}. The renewal character of the BQDs dynamic has been established only recently \cite{paradisi}. 

The similarity between the BQDs and the brain activity dynamics  makes  it plausible to search for a dynamic model accounting for both complex systems.
In this Letter we show that a simplified version of the model of \mbox{Ref.~\cite{katja}} affords significant suggestions on how to realize this important purpose. 
The authors of \mbox{Ref.~\cite{katja}} use a system of coupled \mbox{3-state} stochastic clocks, while we use a system coupled \mbox{2-state} stochastic clocks. 
We shall prove that at the onset of phase synchronization the dynamics of the system has the same properties as the BQDs \cite{bqd,paradisi} and brain 
dynamics \cite{eeg}. They are non-Poisson renewal processes operating in the non-ergodic regime.

\begin{figure}[h]
\includegraphics[angle=-90,width=\linewidth]{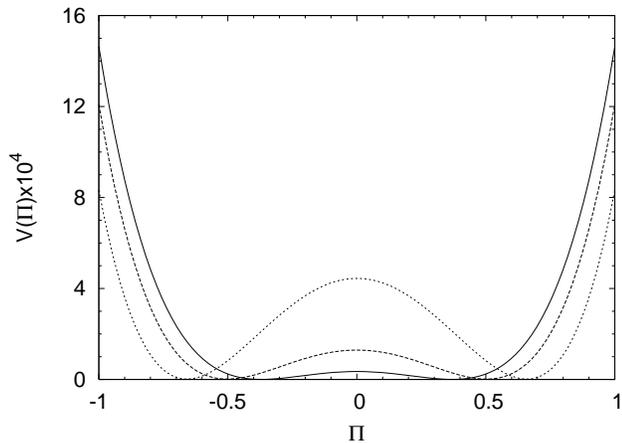}
\caption{The potential $V(\Pi)$, rescaled by a factor $10^{4}$, as a function of $\Pi$ for $g$$=$$0.01$ and different values of the coupling constant $K$. 
$K$$=$$1.05$ for the continuous line, $K$$=$$1.1$ for the dashed line, and $K$$=$$1.2$ for the dot-dashed line.}
\label{figure1}
\end{figure}

We consider a Gibbs ensemble of systems with $N$ \mbox{2-state} stochastic clocks, each of them coupled to $N_{c}$ clocks. We denote by $\mid$$1$$>$ and $\mid$$2$$>$ 
the two states of the clock, corresponding to the phases $\Phi$$=$$0$ and $\Phi$$=$$\pi$, respectively.  
The master equation for a single clock of a system of the Gibbs ensemble is
\begin{equation}
\label{meq_first} 
\left\{ \begin{array}{ll}
\frac{\displaystyle d}{\displaystyle dt} P_{1}  = -g_{12} P_{1} + g_{21} P_{2}  \\
\frac{\displaystyle d}{\displaystyle dt} P_{2}  = -g_{21} P_{2} + g_{12} P_{1} 
\end{array} .\right.
\end{equation}
$P_{1}$ ($P_{2}$) is the probability of finding the clock in the state $\mid$$1$$>$ ($\mid$$2$$>$), and $g_{12}$ ($g_{21}$) is the rate of transitions from
the state $\mid$$1$$>$ ($\mid$$2$$>$) to the state $\mid$$2$$>$ ($\mid$$1$$>$). The transition rates $g_{12}$ and $g_{21}$ are defined by means of the prescription of 
\mbox{Ref.~\cite{katja}}:
\begin{equation}
\label{rate_definition}
g_{12(21)} = g \, \exp \left[K(\pi_{2(1)} - \pi_{1(2)})\right].
\end{equation}
In Eq.~(\ref{rate_definition}), $g$ is the unperturbed transition rate of a single clock, $K$$>$$0$ is the coupling constant and $\pi_{1}$ ($\pi_{2}$) 
is the fraction of the $N_{c}$ coupled clocks that are in the state $\mid$$1$$>$ ($\mid$$2$$>$). The authors of \mbox{Ref.~\cite{katja}} place their clocks in a 
d-dimensional lattice where only the nearest neighbors are coupled. Thus, every clock is coupled to $N_{c}$$=$$2d$ clocks. We adopt, instead, 
an all to all coupling: \mbox{$N_{c}$$=$$N$$-$$1$}. With this choice, when \mbox{$N$$\rightarrow$$\infty$}, the mean 
field approximation \cite{stanley}
\begin{equation}
\label{meanfieldapprox}
\pi_{1(2)}=P_{1(2)}
\end{equation}
is an exact property. Thanks to the mean field approximation of Eq.~(\ref{meanfieldapprox}) and to the normalization 
condition $P_{1}$$+$$P_{2}$$=$$1$, the master equation Eq.~(\ref{meq_first}) reduces to 
\begin{equation}
\label{mastereq_bigpi}
\frac{d\Pi(t)}{dt}\!=\!-2g \cosh(K \Pi) \Pi + 2g \sinh(K \Pi)\!=\!- \frac{\partial V(\Pi)}{\partial \Pi},
\end{equation}
where \mbox{$\Pi$=$P_{1}$$-$$P_{2}$} and \mbox{$\Pi$$\in$$\left[ -1,1 \right]$}. Eq.~(\ref{mastereq_bigpi}) describes the overdamped motion of a particle, 
whose position is $\Pi$, within the potential $V(\Pi)$ \cite{kramers}. Using Eq.~(\ref{mastereq_bigpi}), we find that the potential $V(\Pi)$ is symmetric and the values 
of its minima depend only on the coupling constant $K$. Moreover, we find that there is a critical value $K_{c}$ of the coupling 
parameter $K$ 
\begin{equation}
\label{critical_value}
K = K_{c} = 1
\end{equation}
such that: 1) If \mbox{$K$$\leq$$K_{c}$} the potential $V(\Pi)$ has only one minimum for \mbox{$\Pi_{\textrm{min}}$$=$$0$}; 2) if \mbox{$K$$>$$K_{c}$} 
the potential $V(\Pi)$ is symmetric and has two minima \mbox{$\pm$$\Pi_{\textrm{min}}$} separated by a barrier with the maximum centered in \mbox{$\Pi$$=$$0$}. 
As shown in Fig.~\ref{figure1}, the value $\Pi_{\textrm{min}}$ and the height of the barrier V(0) are increasing function of the coupling constant $K$ \cite{potential}. 
In particular \mbox{$\Pi_{\textrm{min}}$$\rightarrow$$1$} and \mbox{$V(0)$$\rightarrow$$+\infty$} when \mbox{$K$$\rightarrow$$+\infty$}.

\begin{figure}[h]
\includegraphics[angle=-90,width=\linewidth]{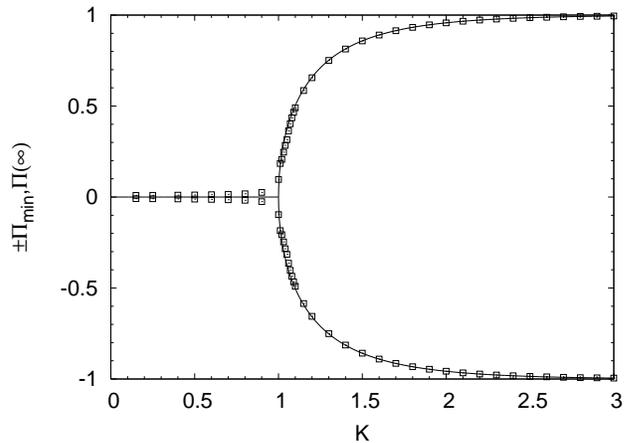}
\caption{The minima $\pm\Pi_{\textrm{min}}$ of the potential $V(\Pi)$, and the asymptotic value $\Pi(\infty)$ as a function of the coupling 
constant $K$. The full line is the theoretical prediction obtained solving $\frac{\partial V(\Pi)}{\partial \Pi}$$=$$0$ and 
$\frac{\partial^{2} V(\Pi)}{\partial \Pi^{2}}$$>$$0$. The squares denote the result of the numerical evaluation of $\Pi(\infty)$ with a Gibbs ensemble of 
systems each with  $N$$=$$10000$ clocks.}
\label{figure2}
\end{figure}

\begin{figure}[h]
\includegraphics[angle=-90,width=\linewidth]{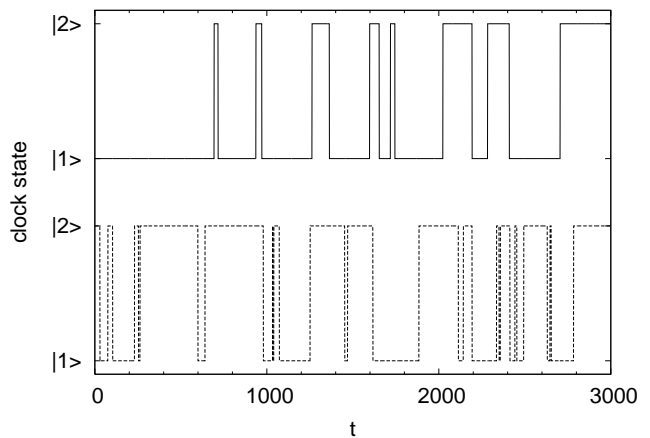}
\caption{The typical time evolution of a single clock of a system in the Gibbs ensemble. The solid (dashed) line refers to the case 
$\Pi(\infty)$$=$$+$$\Pi_{\textrm{min}}$ ($\Pi(\infty)$$=$$-$$\Pi_{\textrm{min}}$). The number of clocks of the single system of the Gibbs 
ensemble is 10000, the unperturbed transition rate is $g$$=$$0.01$ and the coupling constant is $K$$=$$1.05$.} 
\label{figure3}
\end{figure}

The time evolution of the variable $\Pi$ is determined by the minima and maxima of the potential $V(\Pi)$. Thus, two kinds of dynamical evolution are possible: 
1) If \mbox{$K$$\leq$$K_{c}$}, $\Pi(t)$ will reach, after a transient, an asymptotic value $\Pi(\infty)$$=$$0$ not depending on the initial conditions $\Pi(0)$; 2) 
if \mbox{$K$$>$$K_{c}$}, $\Pi(t)$ will reach, after a transient, an asymptotic value $\Pi(\infty)$$=$$+$$\Pi_{\textrm{min}}$$\neq$$0$ 
($\Pi(\infty)$$=$$-$$\Pi_{\textrm{min}}$$\neq$$0$) for an initial condition $\Pi(0)$$>$$0$ ($\Pi(0)$$<$$0$), while the initial condition $\Pi(0)$$=$$0$ 
will result in $\Pi(t)$$=$$0$ $\forall$$t$. 
In Fig.~\ref{figure2} we compare the minima $\pm\Pi_{\textrm{min}}$ of the potential $V(\Pi)$ and the numerical evaluation of $\Pi(\infty)$ 
for different values of the coupling constant $K$. Fig.~\ref{figure2} shows that a phase transition \cite{stanley} occurs at $K$$=$$K_{c}$$=$$1$.
For a  single clock the condition $\Pi(\infty)$$=$$\pm$$\Pi_{\textrm{min}}$$=$$P_{1}$$-$$P_{2}$$\neq$$0$ corresponds to the statistical ``preference''  to be either in 
the state $\mid$$1$$>$ or $\mid$$2$$>$. This is a consequence of the fact that the transition rates $g_{12}$ and $g_{21}$ of 
Eq.~(\ref{rate_definition}) are different if $K$$>$$K_{c}$. Plugging Eq.~(\ref{meanfieldapprox}) into Eq.~(\ref{rate_definition}) and 
allowing $\Pi$ to reach its asymptotic value, we get 
\begin{equation}
\label{different_rates}
g_{12}=g  \exp(-K \Pi(\infty)) \; \neq \; g_{21}=g  \exp(K \Pi(\infty)).
\end{equation}
Fig.~\ref{figure3} confirms the prediction of Eq.~(\ref{different_rates}) showing that if $\Pi(\infty)$$=$$+$$\Pi_{\textrm{min}}$ 
($\Pi(\infty)$$=$$-$$\Pi_{\textrm{min}}$) the single clock spends on average more time  in the state $\mid$$1$$>$ ($\mid$$2$$>$).
The probability density function for the sojourn times in both the preferred and not-preferred state are exponential functions with different mean sojourn times.

Let us now explore the collective behavior of a single system of $N$ clocks under the all to all coupling condition. Following the authors of
\mbox{Ref.~\cite{katja}}, we define the global clock variable $\xi(t)$ as
\begin{equation}
\label{global_clock_1}
\xi(t) = \frac{\sum\limits_{j=1}^{N} \exp (i \; \Phi_{j}(t))}{N} = \frac{N_{1}(t)-N_{2}(t)}{N}.
\end{equation}
The symbol $i$ indicates the imaginary unit, $\Phi_{j}$ is the phase of the j-th clock, $0$ ($\pi$) if the clock is in the state 
$\mid$$1$$>$ ($\mid$$2$$>$), and $N_{1}(t)$ ($N_{2}(t)$) is the number of clocks of the system in state $\mid$$1$$>$
($\mid$$2$$>$) at a time $t$.
When $N$$\rightarrow$$\infty$, the single system becomes a Gibbs ensemble on its own as all the clocks
are identical, and, at the same time, the mean field approximation (Eq.~(\ref{meanfieldapprox})) becomes valid. In this case the master equation for the 
single system is the master equation Eq.~(\ref{meq_first}), where $P_{1}$ ($P_{2}$) is now the probability of finding a clock of the 
system in the state $\mid$$1$$>$ ($\mid$$2$$>$). Finally, from Eq.~(\ref{global_clock_1}), in the limiting case $N$$=$$\infty$, we get 
\begin{equation}
\label{global_clock_2}
\xi(t)=P_{1}(t)-P_{2}(t)=\Pi(t).
\end{equation}
Thus, for a system with infinitely many clocks, the time behavior of the global clock variable $\xi$ is identical to that of the variable $\Pi$ of 
Eq.~(\ref{mastereq_bigpi}): for $K$$>$$K_{c}$ (we exclude the initial unstable condition \mbox{$\xi(0)$$=$$0$})
\begin{equation}\label{global_clock_3}
\xi(\infty)=\pm\Pi_{\textrm{min}}\neq0 \:\:\:\;\;\;\textrm{if} \:\;\xi(0)\gtrless0.
\end{equation}
The condition $\xi(\infty)$$\neq$$0$ of Eq.~(\ref{global_clock_3}) proves that a global phase synchronization  occurs in the 
system at the onset of phase transition ($K$$>$$K_{c}$). Moreover, the time evolution of a single clock of the system is the 
one depicted  by Fig.~\ref{figure3}: state $\mid$$1$$>$ ($\mid$$2$$>$) is statistically preferred if \mbox{$\xi(\infty)$$=$$+$$\Pi_{\textrm{min}}$} 
(\mbox{$\xi(\infty)$$=$$-$$\Pi_{\textrm{min}}$}).

\begin{figure}[h]
\includegraphics[angle=-90,width=\linewidth]{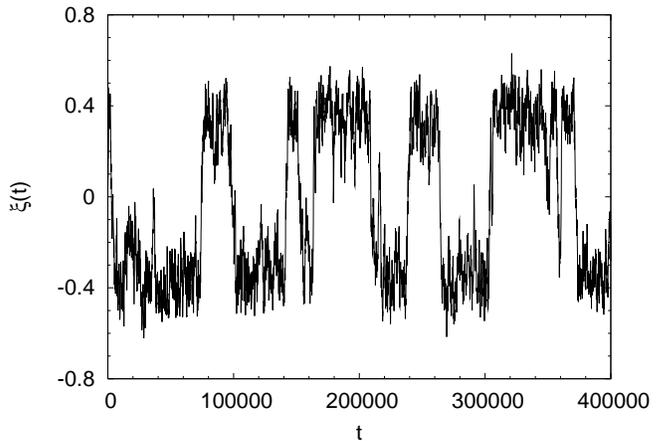}
\caption{The global variable $\xi(t)$ as a function of time, for $K$$=$$1.05$, $g$$=$$0.01$ and a system of 1000 clocks. }
\label{figure4}
\end{figure}

Is the phase synchronization of Eq.~(\ref{global_clock_3}) present in a system with a finite number of clocks? From the definition of transition rate and of 
probability, it follows that
\begin{eqnarray}
\label{rate_meaning} \left\{ \begin{array}{ll}
g_{12}P_{1} = \mathop {\lim } \limits_{N\to \infty} \frac{\displaystyle N_{1\mapsto2}}{\displaystyle N}\\
g_{21}P_{2} = \mathop {\lim } \limits_{N\to \infty} \frac{\displaystyle N_{2\mapsto1}}{\displaystyle N}
\end{array},\right.
\end{eqnarray}
where $N_{1\mapsto2}$ ($N_{2\mapsto1}$) is the number of clocks that undergo a transition from state $\mid$$1$$>$ ($\mid$$2$$>$) to 
state $\mid$$2$$>$ ($\mid$$1$$>$) per unit of time, and $N$ is the number of clocks of the system. Using the law of large numbers \cite{linda}, 
we get that, for any  $N$ fitting the condition $\infty$$>$$N$$\gg$$1$, 
\begin{eqnarray}
\label{lawgrtnum} \left\{ \begin{array}{ll}
\frac{\displaystyle N_{1\mapsto2}}{\displaystyle N} = g_{12}P_{1} + \varepsilon_{12}P_{1}\\
\frac{\displaystyle N_{2\mapsto1}}{\displaystyle N} = g_{21}P_{2} + \varepsilon_{21}P_{2}\\
\end{array},\right.
\end{eqnarray}
where $\epsilon_{12}$ and $\epsilon_{21}$ are fluctuating variables whose intensities are $\propto$$1$$/$$\sqrt(N)$. 
From Eqs.~(\ref{rate_meaning}) and (\ref{lawgrtnum}), we conclude that the master equation of a system with a finite number of clocks is equivalent to 
that of a system with an infinite number of clocks whose transition rates fluctuate:
\begin{equation}
\label{meq_finitesystem} 
\left\{ \begin{array}{ll}
\frac{\displaystyle d}{\displaystyle dt} P_{1}  = -(g_{12}+\varepsilon_{12})P_{1} + (g_{21}+\varepsilon_{21})P_{2}  \\
\frac{\displaystyle d}{\displaystyle dt} P_{2}  = -(g_{21}+\varepsilon_{21})P_{2} + (g_{12}+\varepsilon_{12})P_{1} 
\end{array} .\right.
\end{equation}
If $\infty$$>$$N$$\gg$$1$, we can still consider the mean field approximation of Eq.~(\ref{meanfieldapprox}) to be valid. Using the master equation 
Eq.~(\ref{meq_finitesystem}) and the normalization condition $P_{1}$$+$$P_{2}$$=$$1$, we get for the variable $\Pi$$=$$P_{1}$$-$$P_{2}$ the following 
equation of motion
\begin{equation}
\label{mastereq_bigpi_fluct}
\frac{d\Pi(t)}{dt} = - \frac{\partial V(\Pi)}{\partial \Pi} - \eta(t) \Pi(t) + \theta (t).
\end{equation}
The presence of the fluctuations  $\eta$$=$$\epsilon_{12}+\epsilon_{21}$ and $\theta$$=$$\epsilon_{12}-\epsilon_{21}$ in Eq.~(\ref{mastereq_bigpi_fluct}) 
has the effect of triggering transitions from one well to the other of the potential $V(\Pi)$ of Fig.~\ref{figure1}. Thus, for a system with a finite number of 
clocks the phase synchronization of Eq.~(\ref{global_clock_3}) is not stable. The global clock variable $\xi$ of Eq.(\ref{global_clock_1})  fluctuates 
(for $K$$>$$K_{c}$) between the two minima, $\pm$$\Pi_{\textrm{min}}$, of the potential $V(\Pi)$, as confirmed by Fig.~\ref{figure4}. The single clock follows the 
fluctuations of the global clock variable $\xi$, switching back and forth from the condition where the state $\mid$$1$$>$ is statistically preferred (time evolution 
described by the full line of Fig.~\ref{figure3}) to that where the state $\mid$$2$$>$ is (time evolution described by the dashed line 
of Fig.~\ref{figure3}). 

\begin{figure}[]
\includegraphics[angle=-90,width=\linewidth]{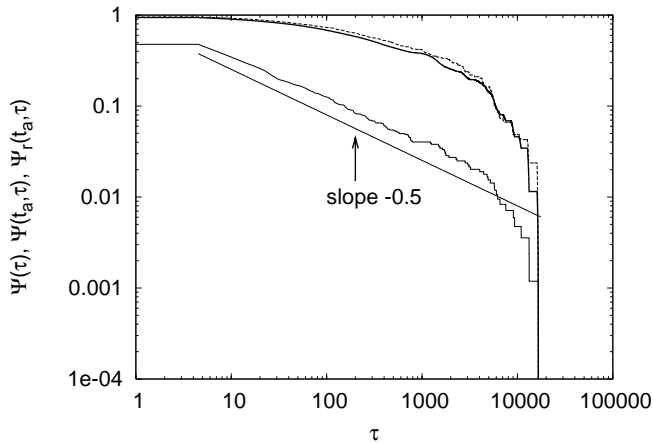}
\caption{The survival probability $\Psi(\tau)$ (full line), the survival probability $\Psi(t_{a},\tau)$ (dashed line) and the survival probability in the renewal 
case $\Psi_{r}(t_{a},\tau)$ (full thick line) as a function of sojourn time $\tau$. Here, we plot  for clarity only the survival probability of age $t_{a}$$=$$500$.}
\label{figure5}
\end{figure}

The probability density functions of the sojourn times in the state $\xi$$>$$0$ or $\xi$$<$$0$ (Fig.~\ref{figure4}) are identical since the 
potential $V(\Pi)$ of Fig.~\ref{figure1} is symmetric. Thus, we denote them with the same symbol $\psi(\tau)$. 
Let us consider a condition where the coupling constant $K$ is close to the critical value \mbox{$K$$-$$K_{c}$$\ll$$1$}. In this case, the height 
of the barrier $V(0)$ dividing the two wells of the 
potential $V(\Pi)$ of Fig.~\ref{figure1} is smaller than or comparable to the intensity of the fluctuations $\eta$ and $\theta$ of Eq.~(\ref{mastereq_bigpi_fluct}). Under 
this condition, we expect \cite{margolin} \mbox{$\psi(\tau)$$\propto$$1/\tau^{1.5}$} for an extended interval of sojourn times. This is exactly what we observe in 
Fig.~\ref{figure5}, where the full line denotes the survival probability $\Psi(\tau)$, namely, the probability of 
observing a sojourn time larger than $\tau$ \cite{survivalprop}. For any fixed value of the unperturbed rate $g$ (Eq.~(\ref{rate_definition})), the height $V(0)$ of 
the barrier dividing the two wells of the potential $V(\Pi)$ (Fig.~\ref{figure1}) increases as the value of the coupling parameter increases. Eventually, the height $V(0)$ 
will  become much larger than the intensity of the fluctuations $\eta$ and $\theta$ (Eq.~(\ref{mastereq_bigpi_fluct})). As a consequence, the power law behavior 
($\psi(\tau)$$\propto$$1/\tau^{1.5}$) of Fig.~\ref{figure5} disappears, the theoretical arguments of  \cite{margolin} loses validity, and an exponential 
behavior emerges, as predicted by the Kramers theory \cite{kramers}. 

Finally, we show that the transition between the states $\xi$$>$$0$ and  $\xi$$<$$0$ 
shown in Fig.~\ref{figure4} is a renewal process. For this purpose we use the aging experiment of Ref.~\cite{paradisi}. We evaluate the survival 
probability $\Psi(t_{a},\tau)$ of age $t_{a}$. This is the probability of observing a sojourn time larger than $\tau$ if the observation starts at a time $t_{a}$ 
after a crossing from $\xi$$>$$0$ to 
$\xi$$<$$0$ or vice versa ($\Psi(0,\tau)$$=$$\Psi(\tau)$). We, then, compare $\Psi(t_{a},\tau)$ with the expected survival probability $\Psi_{r} (t_{a},\tau)$ 
of age $t_{a}$, evaluated according to the renewal theory \cite{paradisi}. If \mbox{$\Psi(t_{a},\tau)$$=$$\Psi_{r}(t_{a},\tau)$ $\forall$$t_{a}$} the process described by 
the survival probability $\Psi(\tau)$ is a renewal process. If $\Psi(\tau)$ is not an exponential function, 
$\Psi(t_{a},\tau)$ yields a slower decay than $\Psi(\tau)$, a condition that is denoted as ``aging'' \cite{paradisi}. 
Fig.~\ref{figure5} shows that the transition between the states $\xi$$>$$0$ and  $\xi$$<$$0$ (Fig.~\ref{figure4}) is a renewal process and that there is aging. With 
increasing values of coupling parameter $K$ the renewal property is not lost, but the aging property is because the survival probability $\Psi(\tau)$ becomes an exponential function \cite{paradisi}. 

In conclusion, we have shown that, for a Gibbs ensemble of systems with $N$$=$$\infty$ coupled 2-state clocks, a phase transition occurs at the critical value 
of coupling parameter $K_{c}$$=$$1$ (Fig.~\ref{figure2}). The phase transition mirrors a statistical ``preference'' for a clock of one of the Gibbs ensemble systems 
to be in one of the two possible states (Eq.(\ref{different_rates})). For a single system with $N$$=$$\infty$ coupled 2-state clocks, the phase 
transition of Fig.~\ref{figure2} signals the onset of a global phase synchronization (Eqs.~(\ref{global_clock_1}), (\ref{global_clock_2}) and (\ref{global_clock_3})). 
If the number of clocks of the system is finite the phase synchronization is not stable. In this case, the variable $\xi$ (Eq.~\ref{global_clock_1}) describing the 
collective motion of the system (Fig.~\ref{figure4}) is characterized by non-Poisson intermittent behavior. At the onset of the phase 
transition (\mbox{$K$$-$$K_{c}$$\ll$$1$}), the zero crossings of the global variable $\xi$ (Fig.~\ref{figure4}) define a series of events with the following properties: 
1) The probability density function of inter events intervals (full line of Fig.~\ref{figure5}) has a non-Poisson non-ergodic character 
(\mbox{$\psi(\tau)$$\propto$$1/\tau^{1.5}$$\Rightarrow$} infinite mean inter event interval). 2) The sequence of inter events intervals satisfy the renewal aging \
condition (Fig.~\ref{figure5}). The properties 1) and 2) are observed properties of the events in both BQDs \cite{bqd, paradisi, margolin} and brain activity \cite{eeg}, 
suggesting that a system consisting of a finite number of coupled \mbox{2-state} clocks may be a good model for the dynamics of both processes. 

The authors thankfully acknowledge Welch and ARO for financial support through Grant no. B-1577 and no. W911NF-05-1-0205, respectively.


\begin{thebibliography}{100}
\bibitem{potential} The arbitrary constant in the defintion of the potential $V(\Pi)$ is chosen to satisfy the condition \mbox{$V(\pm \Pi_{\textrm{min}})$$=$$0$} \mbox{$\forall$$K$$>$$K_{c}$}.
\end{thebibliography}
\end{document}